\documentclass[
 aps,
 prb,
 amsmath,
 amssymb,
 twocolumn
]{revtex4-1}
\usepackage{SIunits}
\usepackage{graphicx}
\usepackage{dcolumn}
\usepackage{bm}

\begin{document}

\title{Large quantum dots with small oscillator strength}
\author{S. Stobbe$^1$}\email[]{ssto@fotonik.dtu.dk}
\author{T. W. Schlereth$^{2,3}$}
\author{S. H\"ofling$^{2,3}$}
\author{A. Forchel$^{2,3}$}
\author{J. M. Hvam$^1$}
\author{P. Lodahl$^1$}\homepage[]{www.fotonik.dtu.dk/quantumphotonics}

\affiliation{
$^{1}$DTU Fotonik, Department of Photonics Engineering, Technical University of Denmark, {\O}rsteds Plads 343, DK-2800 Kgs.~Lyngby, Denmark\\
$^{2}$Technische Physik, Universit\"at W\"urzburg, Am Hubland, D-97074 W\"urzburg, Germany\\
$^{3}$Wilhelm Conrad R\"ontgen-Center for Complex Material Systems (RCCM), Am Hubland, D-97074 W\"urzburg
}

\date{\today}

\begin{abstract}
We have measured the oscillator strength and quantum efficiency of excitons confined in large InGaAs quantum dots by recording the spontaneous emission decay rate while systematically varying the distance between the quantum dots and a semiconductor-air interface. The size of the quantum dots is measured by in-plane transmission electron microscopy and we find average in-plane diameters of 40\,nm. We have calculated the oscillator strength of excitons of that size and predict a very large oscillator strength due to Coulomb effects. This is in stark contrast to the measured oscillator strength, which turns out to be much below the upper limit imposed by the strong confinement model. We attribute these findings to exciton localization in local potential minima arising from alloy intermixing inside the quantum dots.
\end{abstract}

\pacs{78.67.Hc, 42.50.Ct, 78.47.jd}

\maketitle

Enhancement of light-matter interaction is important for improving existing optoelectronic devices such as light-emitting diodes and semiconductor lasers as well as for enabling envisioned devices for quantum information processing. The interaction between light and an emitter can be enhanced by modifying the environment surrounding the emitter, i.e.\ by increasing the optical field using nanophotonic cavities, which can be realized in many geometries such as microdiscs~\cite{Peter2005}, micropillars~\cite{Gerard1998} or photonic crystal cavities~\cite{Laucht2009}. Cavity enhancement works by increasing the local density of optical states, which describes the number of vacuum modes that an emitter can radiate into. Another approach to enhance the light-matter interaction is to modify the emitter, i.e.\ to tailor the matter-part. The relevant figure-of-merit is the oscillator strength, which is a dimensionless quantity defined as the ratio between the radiative decay rate of the emitter in a homogeneous medium and the emission rate of a classical harmonic oscillator.

Self-assembled quantum dots are particularly interesting light-emitters because their oscillator strength is typically one order of magnitude larger than that of atoms~\cite{Hens2008}. Furthermore, as first pointed out by Hanamura~\cite{Hanamura1988}, the oscillator strength of excitons in a large quantum dot is proportional to the volume of the quantum dot: In this case Coulomb effects dominate the electron-hole confinement and the exciton acquires the sum of oscillator strengths of all lattice sites that it spans. This giant oscillator strength effect arises in the weak confinement regime, i.e.\ when the confinement is so weak that the energy level spacing is smaller than the Coulomb energy.

For small quantum dots the level spacing is much larger than the Coulomb energy and the exciton state can be described by a product of an electron state and a hole state, which are mutually independent. This is known as the strong confinement regime, which is the relevant regime for the majority of contemporary experiments. In the strong confinement approximation the oscillator strength is proportional to the square of the electron and hole envelope function overlap, which sets an upper limit to the achievable oscillator strength because the overlap cannot exceed unity~\cite{Stobbe2009}.

It was predicted by Andreani et al.~\cite{Andreani1999} in 1999 that large quantum dots are essential to reach the strong coupling regime of light-matter interaction and indeed the vacuum Rabi splitting signature of strong coupling has been observed with large GaAs quantum dots~\cite{Peter2005}. Also, a very high oscillator strength of large GaAs quantum dots has been reported~\cite{Hours2005}. Here we report on direct measurements of the oscillator strength of large In$_\mathrm{0.3}$Ga$_\mathrm{0.7}$As quantum and surprisingly observe no enhancement of the oscillator strength beyond the strong confinement limit.

The simplest way to measure the oscillator strength of a quantum dot would be to extract it from a measurement of the radiative decay rate in a homogeneous medium. However, the radiative decay rate is not obtained directly by time-resolved spectroscopy, which extracts the total decay rate, i.e.\ the sum of radiative and nonradiative decay rates. The contribution from nonradiative decay processes was recently found to be significant for small quantum dots~\cite{Johansen2008,Stobbe2009} and the nonradiative decay rate was not measured in any of the previous experimental studies on large quantum dots.

Here we use a recently developed experimental method~\cite{Johansen2008,Stobbe2009} to accurately measure the radiative and nonradiative decay rates and thereby extract the oscillator strength and quantum efficiency of large InGaAs quantum dots. We find that the experimentally determined oscillator strength is below the upper limit of the strong confinement model and that the relatively fast decay rates originate from a quantum efficiency of only 33\%, i.e.\ the decay rate is dominated by nonradiative decay. Our results show that the effective confinement potential in quantum dots can be much smaller than the quantum dot size presumably due to local variations in strain and chemical composition of the quantum dots.

\begin{figure}
\includegraphics[width=\columnwidth]{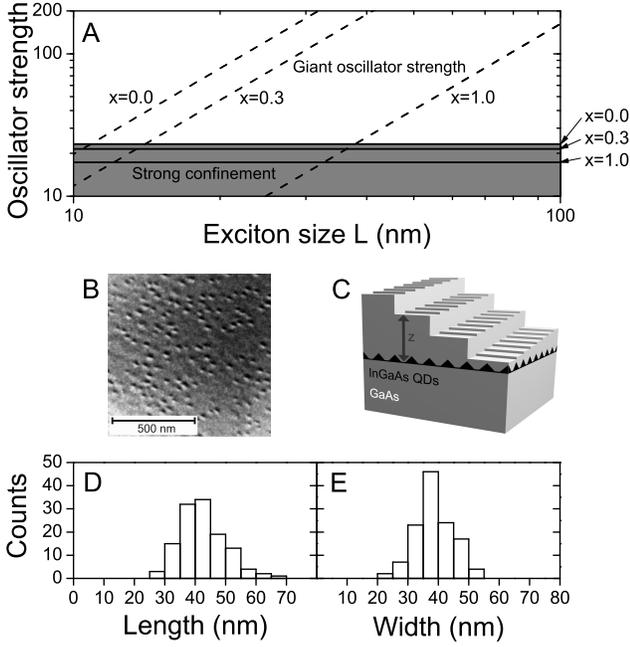}
\caption{\label{OscillatorStrengths_SC_vs_WC} (A) Calculated oscillator strength for In$_x$Ga$_{1-x}$As quantum dots with $x=0$, $x=0.3$, and $x=1$. In the weak confinement model (dashed curves) the oscillator strength increases quadratically with exciton size, which is known as the giant oscillator strength effect. In strong confinement there is an upper bound (solid curves) to the oscillator strength at $f=23.2$. The gray area indicates the regime of strong confinement for various mole fractions $x$. (B) In-plane transmission electron micrograph of the overgrown In$_{0.3}$Ga$_{0.7}$As quantum dots. (C) Schematic illustration of the sample investigated. (D) Distribution of quantum dot lengths along the major axis. (E) Distribution of quantum dot widths along the minor axis. The average size along the major (minor) axis is found to be 42\,nm (38\,nm).}
\end{figure}

The maximum attainable oscillator strength of In$_{x}$Ga$_{1-x}$As quantum dots in the strong confinement model is given by~\cite{Johansen2008,Stobbe2009}
\begin{equation}
f_\mathrm{SC,max} (\omega,x) = \frac{E_p(x)}{\hbar\omega},
\end{equation}
with the Kane energy $E_p(x) = (28.8-7.3x)\:\mathrm{eV}$ and $\hbar\omega$ is the exciton transition energy. In the weak confinement regime the decay rate can be calculated using Wigner-Weisskopf theory for excitons confined in a parabolic in-plane potential perpendicular to the growth direction and a hard-wall potential along the growth direction. For the lowest energy transition the oscillator strength is
\begin{equation}
f_\mathrm{WC} (\omega,x) = \frac{2 E_p(x)}{\hbar\omega}\left(\frac{L}{a_0}\right)^2,\label{eq:OS_WC}
\end{equation}
where $L$ is the diameter of the center-of-mass wave function in the plane perpendicular to the growth direction~\cite{Sugawara1995}, which we define as four standard deviations of the Gaussian wave function and $a_0$ is the exciton Bohr radius, which is defined as $a_0=\frac{4\pi\hbar^2\epsilon_0\epsilon_r}{q^2 m_0 m}$, where $\epsilon_0$ is the vacuum permittivity, $\epsilon_r$ is the relative dielectric constant of GaAs, $q$ is the electron charge, $m_0$ is the electron rest mass, and $m=\frac{m_\mathrm{e} m_\mathrm{hh}}{m_\mathrm{e}+m_\mathrm{hh}}$ is the reduced effective mass of the exciton composed of an electron with effective mass $m_\mathrm{e}$ and a heavy-hole with effective mass $m_\mathrm{hh}$. The effective masses depend on the indium mole fraction of the quantum dot and the heavy-hole effective mass is modified by strain. For In$_{0.3}$Ga$_{0.7}$As we obtain $a_0 = 19.2\:\mathrm{nm}$ using parameters from Ref.~\cite{Stobbe2009}, where we are considering only the heavy-hole mass in the plane perpendicular to the growth direction.

The oscillator strength calculated in the weak confinement model is plotted in Fig.~\ref{OscillatorStrengths_SC_vs_WC}(A) along with the fundamental oscillator strength limit of the strong confinement model for various indium mole fractions. Thus, by comparing measured and calculated oscillator strengths the proper confinement model can be identified and such an analysis is presented in the following section.

Upon measurement of the radiative decay rate of a quantum dot in a homogeneous medium $\Gamma_\mathrm{rad}^\mathrm{hom}(\omega)$ and the nonradiative decay rate $\Gamma_\mathrm{nrad}(\omega)$, the oscillator strength can be obtained directly from the equation
\begin{equation}
f(\omega) = \frac{6 \pi m_0 \epsilon_0 c_0^3}{n(\omega) q^2 \omega^2} \Gamma_\mathrm{rad}^\mathrm{hom}(\omega),
\end{equation}
where $c_0$ is the speed of light in vacuum and $n(\omega)$ is the index of refraction of GaAs, which depends on $\omega$ as well as the temperature~\cite{Gehrsitz2000}.
The quantum efficiency is defined as
\begin{equation}
QE(\omega) = \frac{\Gamma_\mathrm{rad}^\mathrm{hom}(\omega)}{\Gamma_\mathrm{rad}^\mathrm{hom}(\omega) + \Gamma_\mathrm{nrad}(\omega)}.
\end{equation}

The starting point of the experimental investigation was a semiconductor wafer grown by molecular beam epitaxy. First a 50$\:\mathrm{nm}$ AlAs sacrificial layer for an optional epitaxial lift-off process was grown on a GaAs substrate. This was followed by 1038$\:\mathrm{nm}$ of GaAs, a layer of large InGaAs quantum dots with a nominal indium content of 30\%, and finally a 445$\:\mathrm{nm}$ GaAs capping layer. The quantum dot growth procedure was similar to a method discussed in the literature~\cite{Loeffler2006} but differed by employing growth interruptions of in total 15\,s. The quantum dots had a density of approximately $150\:\mathrm{\micro m}^{-2}$ and were slightly elliptically shaped with typical major and minor axis diameters of 42$\:\mathrm{nm}$ and 38$\:\mathrm{nm}$ respectively, as determined from in-plane transmission electron microscopy on the overgrown sample~\cite{Werner2000}, cf.\ Fig.~\ref{OscillatorStrengths_SC_vs_WC}(B), (D), and (E). Thus, from Eq.~(\ref{eq:OS_WC}) and cf.\ Fig.~\ref{OscillatorStrengths_SC_vs_WC}(A) we would expect $f\gtrsim 100$ for these quantum dots.

The capping layer was processed into 32 terraces, which measured 200~$\micro\mathrm{m}$ by 500~$\micro\mathrm{m}$ thus constituting 32 different distances from the quantum dot layer to the interface, as shown in Fig.~\ref{OscillatorStrengths_SC_vs_WC}(C). For each distance we have performed time-resolved measurements of quantum dot ensembles at a temperature of 19~K and for one distance we performed time-resolved microphotoluminescence spectroscopy on single quantum dots at 10~K. All measurements were acquired using pulsed excitation from a Ti:sapphire laser at a wavelength of 860$\:\mathrm{nm}$, i.e.\ in the wetting layer of the quantum dots. Further details on the sample preparation, measurement setup, and theoretical approach can be found in Ref.~\cite{Stobbe2009}. In the following we discuss the experimental results.

\begin{figure}
\includegraphics[width=\columnwidth]{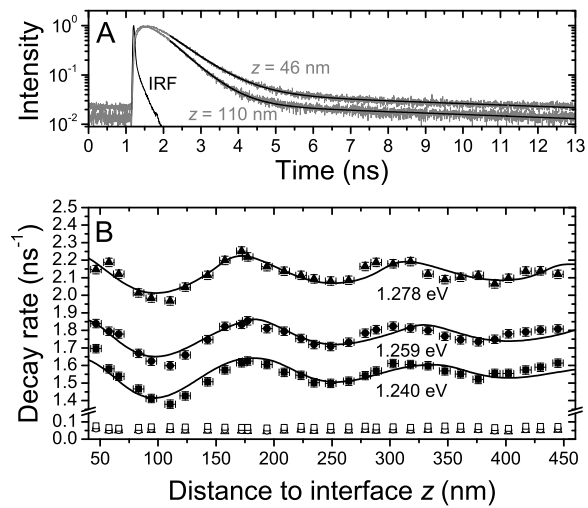}
\caption{\label{M3309_Decays_and_oscillations} (A) Characteristic decay curves (gray) of quantum dots obtained at two different distances to the interface at an emission energy of 1.240$\:\mathrm{eV}$ with biexponential fits (black). The curve labeled IRF is the instrument response function. (B) Fast (closed symbols) and slow (open symbols) decay rates obtained from the biexponential fits for various distances to the interface at an emission energy of 1.240$\:\mathrm{eV}$ (squares), 1.259$\:\mathrm{eV}$ (circles), and 1.278$\:\mathrm{eV}$ (triangles).}
\end{figure}

In Fig.~\ref{M3309_Decays_and_oscillations}(A) we show two characteristic decay curves for two different distances to the interface. The decays are markedly different due to the different values of the local density of optical states. We fit the curves by bi-exponential decay functions convoluted with the instrument response function and find very good agreement. We have measured and fitted the decay curves for 32 distances to the interface and at three different emission energies of the quantum dot ensemble. For the two distances closest to the interface ($z=12$ and $z=28\:\mathrm{nm}$) we observe no photoluminescence, which presumably is due to tunneling of carriers to the surface followed by nonradiative recombination. The extracted fast and slow components for the remaining 30 distances are shown in Fig.~\ref{M3309_Decays_and_oscillations}(B). The slow decay rate is independent of distance to the interface, which shows that it is dominated by nonradiative decay~\cite{Johansen2008DarkExcitons}. In the remainder of this article we discuss only the fast decay rate, which exhibits a characteristic oscillatory behavior as a function of distance.

We have calculated the local density of optical states for a dipole source oriented parallel to the interface. The calculation is exact and takes all layers above and below the quantum dots into account. We have fitted the decay rate to the local density of optical states as a function of distance to the interface and we find very good agreement as is evident from the fits shown in Fig.~\ref{M3309_Decays_and_oscillations}(B).

\begin{figure}
\includegraphics[width=\columnwidth]{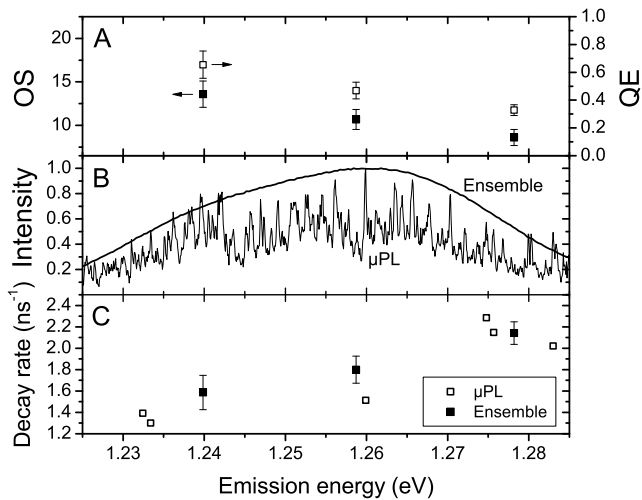}
\caption{\label{FreqDepOfOSandQE} (A) Oscillator strength (OS, solid symbols, left axis) and quantum efficiency (QE, open symbols, right axis) for different emission energies. (B) Normalized photoluminescence spectrum obtained by ensemble and microphotoluminescence ($\mathrm{\micro}$PL) measurements at $z=445\:\mathrm{nm}$. (C) Total homogenous medium decay rate (solid symbols) extracted from the analysis of Fig.~\ref{M3309_Decays_and_oscillations} compared to the microphotoluminescence measurements of the total decay rate at $z=445\:\mathrm{nm}$ (open symbols)).}
\end{figure}

The fit to the local density of optical states for each emission energy has two free parameters: the radiative decay rate and the nonradiative decay rate in a homogeneous medium. From this we obtain readily the oscillator strength and quantum efficiency. The result of this analysis is shown in Fig.~\ref{FreqDepOfOSandQE}(A). We observe a frequency dependence, which is similar to that observed for small quantum dots~\cite{Stobbe2009}, i.e.\ both the oscillator strength and the quantum efficiency decrease with increasing energy. However, in the present case the nonradiative decay rate is much larger leading to quantum efficiencies between $(65 \pm 10)\%$ and $(33 \pm 4)\%$. These results show that the oscillator strength is not particularly large for these large quantum dots. In fact the oscillator strength is smaller than the limit imposed by of the strong confinement model, cf.\ Fig.~\ref{OscillatorStrengths_SC_vs_WC}(A). This shows that no giant oscillator strength effect is found for these large InGaAs quantum dots and that they can be described fully within the strong confinement model. Interestingly, the oscillator strength varies between $13.6 \pm 1.5$ and $8.6 \pm 0.9$ which is comparable to or even below the values found for ordinary quantum dots~\cite{Johansen2008}.

It could be conjectured that only few of the large quantum dots possess a large oscillator strength and if their relative density is small they can be revealed only by microphotoluminescence experiments. We performed such an experiment and measured the spectrum shown in Fig.~\ref{FreqDepOfOSandQE}(B), which has been obtained at an excitation power below saturation of the excitons. Single-quantum dot lines are observed and the result of time-resolved measurements on a selection of the peaks are presented in Fig.~\ref{FreqDepOfOSandQE}(C). The decay rates agree very well with the total decay rates extracted from the ensemble measurements, which are also shown in Fig.~\ref{FreqDepOfOSandQE}(C). This indicates that the measurements on ensembles probe the single-quantum dot properties with the additional benefit of statistical averaging and an improved signal-to-noise ratio.

Our results show that the effective size of the excitons is much smaller than the size of the quantum dots as it appears from transmission electron microscopy. This means that the confinement potential is significantly smaller than the quantum dot size and we conjecture that the actual confinement potentials are defined by fluctuations in the local indium/gallium mole fraction. Such fluctuations have been observed in high-resolution transmission electron microscopy studies of small InAs quantum dots~\cite{Wang2006} and our results suggest that they may be of large significance for the optical properties of large quantum dots. In this picture the excitons are confined in potential minima surrounded by barriers and other local potential minima. This could imply large nonradiative decay rates due to tunneling to other local minima, which is exactly what we observe. Furthermore, tunneling of carriers into the potential minimum responsible for recombination from other nonradiative minima could lead to filling effects in the decay curves, i.e.\ decay curves, which appear flat on the top due the re-excitation provided by the tunneling, which is also a feature that is apparent in our measurements, cf.\ Fig.~\ref{M3309_Decays_and_oscillations}(A). We note that similar deviations between the size of the device and the effective potentials have been reported for other mesoscopic electronic systems, such as unintentional quantum dots in high-mobility two-dimensional electron gases~\cite{Ebbecke2005} and carbon nanotubes, where the effective quantum dot size is given by the distance between the electrical contacts rather than the length of the nanotubes~\cite{Sapmaz2005}.

In conclusion we have measured the oscillator strength and quantum efficiency of large InGaAs quantum dot. We find that the decay dynamics is dominated by nonradiative decay processes and that the oscillator strength is comparable to or even smaller than the values reported for small quantum dots. We conclude that the actual confinement potential in these quantum dots is much smaller than the quantum dot size obtained from transmission electron microscopy on the overgrown sample. Our results emphasize the importance of addressing the quantum efficiency of quantum dot emitters because the prevalent assumption of a quantum efficiency of unity is generally not valid and can lead to wrong conclusions.

\begin{acknowledgments}
We would like to acknowledge P.~T.~Kris\-ten\-sen and J.~E.~Mor\-ten\-sen for valuable theoretical discussions and Q.~Wang for assistance with the microphotoluminescence measurements. We gratefully acknowledge the Danish Research Agency (Projects No. FNU 272-05-0083, No. FNU
272-06-0138, and No. FTP 274-07-0459) and the European Commission (Project QPhoton) for financial
support.
\end{acknowledgments}

\end{document}